# Enhancements of the superconducting transition temperature within the two-band model

Annette Bussmann-Holder, Roman Micnas* and Alan R. Bishop**

Max-Planck-Institute for Solid State Research, Heisenbergstr. 1, D-70569 Stuttgart, Germany

*Institute of Physics, A. Mickiewicz University, 85 Umultowska St., 61-614 Poznan, Poland

**Los Alamos National Laboratory, Theoretical Division, Los Alamos, NM87545, USA

The two-band model as introduced by Suhl, Matthias and Walker [Phys. Rev. Lett. **3**, 551 (1959)] accounts for multiple energy bands in the vicinity of the Fermi energy which could contribute to electron pairing in superconducting systems. Here, extensions of this model are investigated wherein the effects of coupled superconducting order parameters with different symmetries and the presence of strong electron-lattice coupling on the superconducting transition temperature $T_c$ are studied. Substantial enhancements of $T_c$ are obtained from both effects.

Shortly after the BCS theory [1] for superconductivity was presented, Suhl, Matthias and Walker proposed extensions of this theory [2] to account for more complex electronic band structures. Their assumption that pairing might occur in various energy bands that are located in the vicinity of the Fermi energy implied that interband interactions between those bands take place in order to assure a homogeneous superconducting state. Interestingly they observed that a two-order parameter scenario leads to an enhancement of the superconducting transition temperature as compared to a single band model. The two-band model has since then been explored more deeply by various groups [3-6] and has also been invoked recently to explain high temperature superconductivity in copper oxides [7-12] and $MgB_2$ [13-15]. In this Letter we study new extensions of the two-band model. We investigate the effect of the coexistence of a dynamic polaronic lattice distortion with superconductivity on the superconducting transition temperature $T_c$. In addition we admit for anisotropic pairing and the mixing of d- and s-wave superconducting order parameters. In all our calculations we start from the assumption that the pairing interactions within the two bands considered are too weak to induce superconductivity separately. Thus we are able to investigate the effect of the interband interactions on anisotropic superconductivity and also show how much a polaronic distortion can influence superconductivity.

The two band Hamiltonian we consider, in already condensed form, reads:

$$H = H_0 + H_1 + H_2 + H_{12} \qquad (1)$$

$$H_0 = \sum_{k_1 s} x_{k_1} c^+_{k_1 s} c_{k_1 s} + \sum_{k_2 s} x_{k_2} d^+_{k_2 s} d_{k_2 s} \qquad (1a)$$

$$H_1 = -\sum_{k_1 k_1' q} V_1(k_1, k_1') c^+_{k_1+q/2\uparrow} c^+_{-k_1+q/2\downarrow} c_{-k_1'+q/2\downarrow} c_{k_1'+q/2\uparrow} \qquad (1b)$$

$$H_2 = -\sum_{k_2 k_2' q} V_2(k_2, k_2') d^+_{k_2+q/2\uparrow} d^+_{-k_2+q/2\downarrow} d_{-k_2'+q/2\downarrow} d_{k_2'+q/2\uparrow} \qquad (1c)$$

$$H_{12} = -\sum_{k_1 k_2 q} V_{12}(k_1, k_2) \{ c^+_{k_1+q/2\uparrow} c^+_{-k_1+q/2\downarrow} d_{-k_2+q/2\downarrow} d_{k_2+q/2\uparrow} + h.c. \}, \qquad (1d)$$

where $H_0$ is the kinetic energy of the bands $i=1,2$ with $x_{k_i} = e_i + e_{k_i} - m$. Here $e_i$ denotes the position of the c and d band with creation and annihilation operators $c^+, c, d^+, d$, respectively, and $m$ is the chemical potential. The pairing potentials $V_i(k_i, k_i')$ are assumed to be represented in factorized form as $V_i(k_i, k_i') = V_i j_{k_i} y_{k_i'}$ where $j_{k_i}, y_{k_i}$ are cubic harmonics for anisotropic pairing which yields for two dimensions and on-site pairing: $j_{k_i} = 1, y_{k_i} = 1$, extended s-wave: $j_{k_i} = \cos k_x a + \cos k_y b = g_{k_i}$ and d-wave: $j_{k_i} = \cos k_x a - \cos k_y b = h_{k_i}$, where *a*, *b* are the lattice constants along *x* and *y* directions. By performing a BCS meanfield analysis of Equs. 1 those transform to:

$$H_{red} = \sum_{k_1 s} x_{k_1} c^+_{k_1 s} c_{k_1 s} + \sum_{k_2 s} x_{k_2} d^+_{k_2 s} d_{k_2 s} + \overline{H}_1 + \overline{H}_2 + \overline{H}_{12} \tag{2}$$

$$\overline{H}_i = -\sum_{k'_i} \left[ \Delta^*_{k'_i} c^+_{k'_i\uparrow} c^+_{-k'_i\downarrow} + \Delta_{k'_i} c_{-k'_i\downarrow} c_{k'_i\uparrow} \right] + \sum_{k_i, k'_i} V_i(k_i, k'_i) < c^+_{k_i\uparrow} c^+_{-k_i\downarrow} >< c_{-k'_i\downarrow} c_{k'_i\uparrow} > ,\ i = 1,2 \tag{2a}$$

and for $i$=2 $c$ is replaced by $d$.

$$\overline{H}_{12} = -\sum_{k_1,k_2} [V_{12}(k_1,k_2) < c^+_{k_1\uparrow} c^+_{-k_1\downarrow} > d_{-k_2\downarrow} d_{k_2\uparrow} + V_{12}(k_1,k_2) < d_{-k_2\downarrow} d_{k_2\uparrow} > c^+_{k_1\uparrow} c^+_{-k_1\downarrow} +$$
$$V^*_{12}(k_1,k_2) d^+_{k_2\uparrow} d^+_{-k_2\downarrow} < c_{-k_1\downarrow} c_{k_1\uparrow} > + V^*_{12}(k_1,k_2) c_{-k_1\downarrow} c_{k_1\uparrow} < d^+_{k_2\uparrow} d^+_{-k_2\downarrow} > -$$
$$V_{12}(k_1,k_2) < c^+_{k_1\uparrow} c^+_{-k_1\downarrow} >< d_{-k_2\downarrow} d_{k_2\uparrow} > - V^*_{12}(k_1,k_2) < c_{-k_1\downarrow} c_{k_1\uparrow} >< d^+_{k_2\uparrow} d^+_{-k_2\downarrow} >]$$

(2b)

Here we assume that $< c^+_{k_1+q/2\uparrow} c^+_{-k_1+q/2\downarrow} > = < c^+_{k_1\uparrow} c^+_{-k_1\downarrow} > d_{q,0}$ and equivalently for the $d$ operators. In addition the following definitions are introduced: $\Delta_{k'_i} = \sum_{k_i} V_i(k_i, k'_i) < c^+_{k_i\uparrow} c^+_{-k_i\downarrow} >$ together with:

$A_{k_1} = \sum_{k_2} V_{12}(k_1,k_2) < d^+_{k_2\uparrow} d^+_{-k_2\downarrow} >$, $B_{k_1} = \sum_{k_2} V_{12}(k_1,k_2) < c^+_{k_2\uparrow} c^+_{-k_2\downarrow} >$ and $V^*_{12}=V_{12}$.

Applying the standard technique we obtain:

$$< c^+_{k_1\uparrow} c^+_{-k_1\downarrow} > = \frac{\overline{\Delta}_{k_1}}{2E_{k_1}} \tanh[\frac{bE_{k_1}}{2}] = \overline{\Delta}_{k_1} \Phi_{k_1} \tag{3a}$$

$$< d^+_{k_2\uparrow} d^+_{-k_2\downarrow} > = \frac{\overline{\Delta}_{k_2}}{2E_{k_2}} \tanh[\frac{bE_{k_2}}{2}] = \overline{\Delta}_{k_2} \Phi_{k_2} \tag{3b}$$

with $E^2_{k_1} = x^2_{k_1} + |\overline{\Delta}_{k_1}|^2, \overline{\Delta}_{k_1} = \Delta_{k_1} + A_{k_1}$ and $E^2_{k_2} = x^2_{k_2} + |\overline{\Delta}_{k_2}|^2, \overline{\Delta}_{k_2} = \Delta_{k_2} + B_{k_2}$. From this we obtain the selfconsistent set of equations:

$$\overline{\Delta}_{k_1} = \sum_{k'_1} V_1(k_1,k'_1) \overline{\Delta}_{k'_1} \Phi_{k'_1} + \sum_{k_2} V_{1,2}(k_1,k_2) \overline{\Delta}_{k_2} \Phi_{k_2} \tag{4a}$$

$$\overline{\Delta}_{k_2} = \sum_{k'_2} V_2(k_2,k'_2) \overline{\Delta}_{k'_2} \Phi_{k'_2} + \sum_{k_1} V_{1,2}(k_1,k_2) \overline{\Delta}_{k_1} \Phi_{k_1} \tag{4b}$$

from which the temperature dependences of the gaps and the superconducting transition temperature have to be determined. If the interactions $V$ are constants, the resulting gaps are momentum independent. A more interesting case is obtained by assuming the following general momentum dependence of the intraband interactions: $V_i = g_0^{(i)} + g_g^{(i)} g_k g_{k'} + g_h^{(i)} h_k h_{k'}$ where the first term yields onsite pairing, the second extended s-wave pairing, and the last term d-wave pairing. In our calculation we assume that $V_1$ is proportional to $g_0$ while $V_2$ is either determined by $g_0$ or by $g_h$. In addition the two bands considered are 1-dimensional in the case of the $c$ bands while the $d$-related band is 2-dimensional with the following dispersion: $e_{k_2} = -2t(\cos k_x a + \cos k_y b)$. As already outlined earlier, throughout this paper we choose our values for the intraband interactions such that both bands separately do not exhibit superconductivity. Specifically, $V_1$=$V_2$=0.01, where $V_1 = \tilde{V}_1 N_s, V_2 = \tilde{V}_2 N_d$. Within this scenario the selfconsistent set of equations is solved numerically as a function of $V_{12} = \tilde{V}_{12} \sqrt{N_s N_d}$ , where $N_s$, $N_d$ are the density-of-states of bands *1*, *2*, respectively. The results are shown in figure 1 where both cases $V_2$ ~$g_0$ and $V_2$~ $g_h$ are considered. In both cases small values of $V_{12}$ are sufficient to induce superconductivity. With increasing $V_{12}$ dramatic enhancements of $T_c$ are obtained which easily exceed 100K. Interestingly the d-wave component in the two component systems has an additional $T_c$-increasing factor which increases with increasing interband coupling strength.

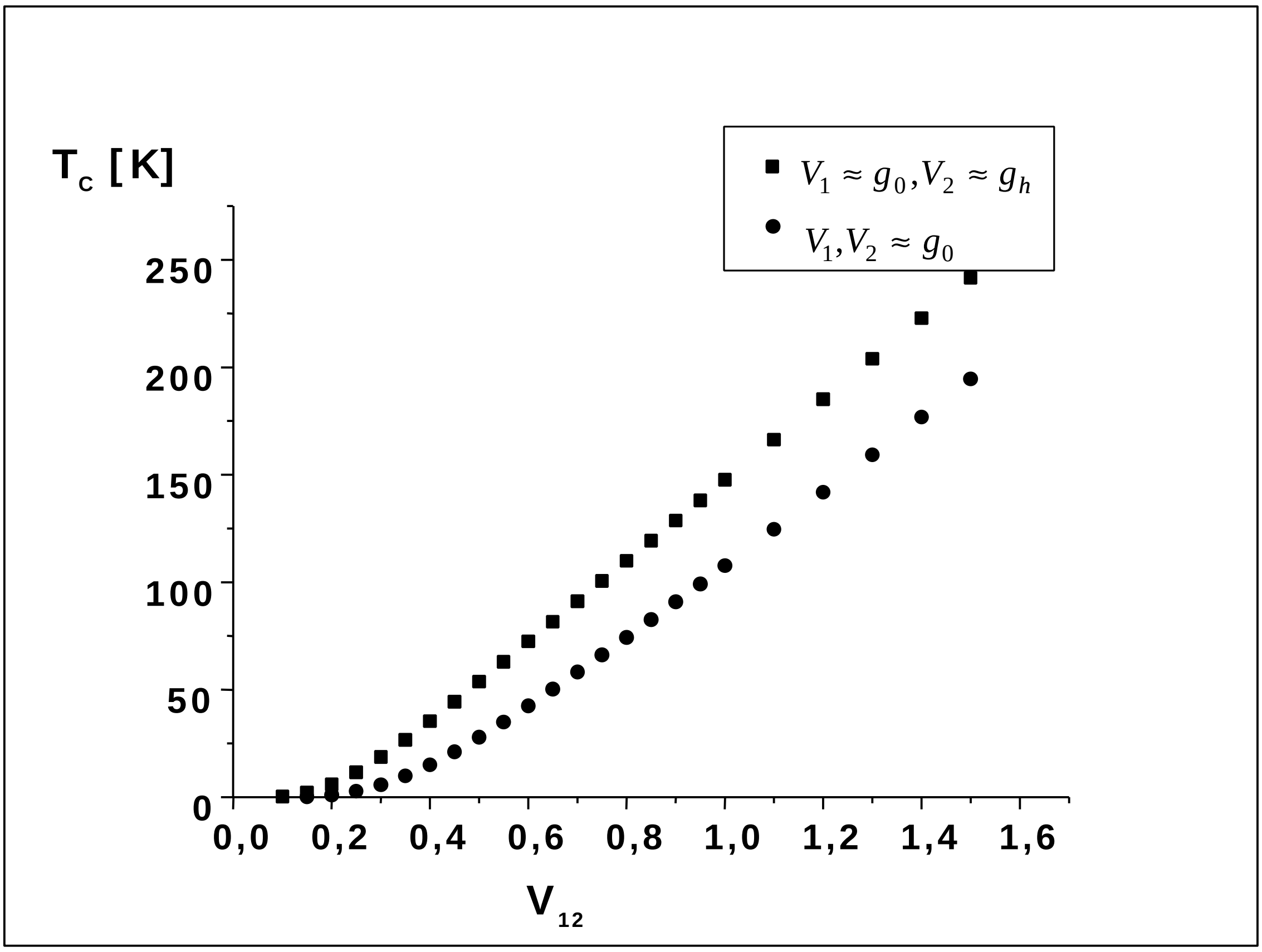

Figure 1: The dependence of the superconducting transition temperature on the interband coupling $V_{12}$ for the case of both, $V_1$ and $V_2$ ~$g_0$ (circles) and the case where $V_1$~$g_0$, $V_2$~ $g_h$ (squares).

This finding demonstrates that a mixed order parameter symmetry favours superconductivity, as opposed to two onsite pairing interactions. It has to be mentioned here, that a mixed order parameter symmetry is not possible on a cubic lattice, but that an orthorhombic distortion has to be considered. By choosing a 10% orthorhombicity an additional small enhancement of $T_c$ as compared to cubic symmetry is obtained, but the general results are overall not affected.

The related superconducting energy gaps are shown in figure 2 with $V_{12}$=0.5. Here again the effect of onsite couplings only depresses the gaps as compared to s/d –wave coupled gaps, and in addition a strong anisotropy of the two gaps is observed within the mixed order parameter system. In the insert the ratio of the gaps with respect to $T_c$ is shown as a function of $T_c$ for the mixed order parameter case only. Interestingly the s-wave gap ratio is close to the BCS ratio, slightly increasing with increasing $T_c$. The corresponding ratio of the d-wave gap is enhanced, as compared to a one band approach, and remains nearly constant as a function of $T_c$ with a slight decrease at small $T_c$'s. The gap versus temperature behaviour is comparable to the conventional two-band model and follows a BCS type temperature dependence.

Finally we have investigated the question of how the coexistence of dynamic polaronic lattice distortion with superconductivity influences $T_c$. We start with the assumption that for temperatures $T >> T_c$ a strong coupling of the one-dimensional electronic band to phonons with momentum $q$-dependent energy $\hbar w$ takes place. This corresponds to modifying the first part of equation 1a as:

$$\overline{H}_0 = \sum_{k_1 s} x_{k_1} c^+_{k_1 s} c_{k_1 s} + \sum_q \hbar w_q b^+_q b_q + \frac{1}{\sqrt{2N}} \sum_{q,s,k_1} g(q) c^+_{k_1+qs} c_{k_1 s} (b_q + b^+_{-q}) \quad (5)$$

Here $b^+$, $b$ are phonon creation and annihilation operators and $g(q)$ is the electron-phonon coupling. Following the procedure of Ref.16 the $k_1$-related electronic energies are renormalized by the electron phonon coupling as: $\tilde{H}_0 = \sum_{k_1 s} (x_{k_1} - \Delta^*) c^+_{k_1 s} c_{k_1 s}$ with $\Delta^* = \frac{1}{2N} \sum_q (\hbar w_q)^{-1} |g(q)|^2$. The transformation to small polarons yields an additional exponential reduction in the hopping integrals which is not relevant for the one dimensional band considered here, but has to be included if similar effects were discussed for the two dimensional band. The polaronic induced density-density attraction has been absorbed in the coupling constant $V_1$. The q-dependence of the electron-phonon coupling together with that of the level shift have been treated here as integrated averaged quantities but they are explicitly taken into account in ref. 18. Including these modifications of the one dimensional electronic band in the calculation of $T_c$, and considering

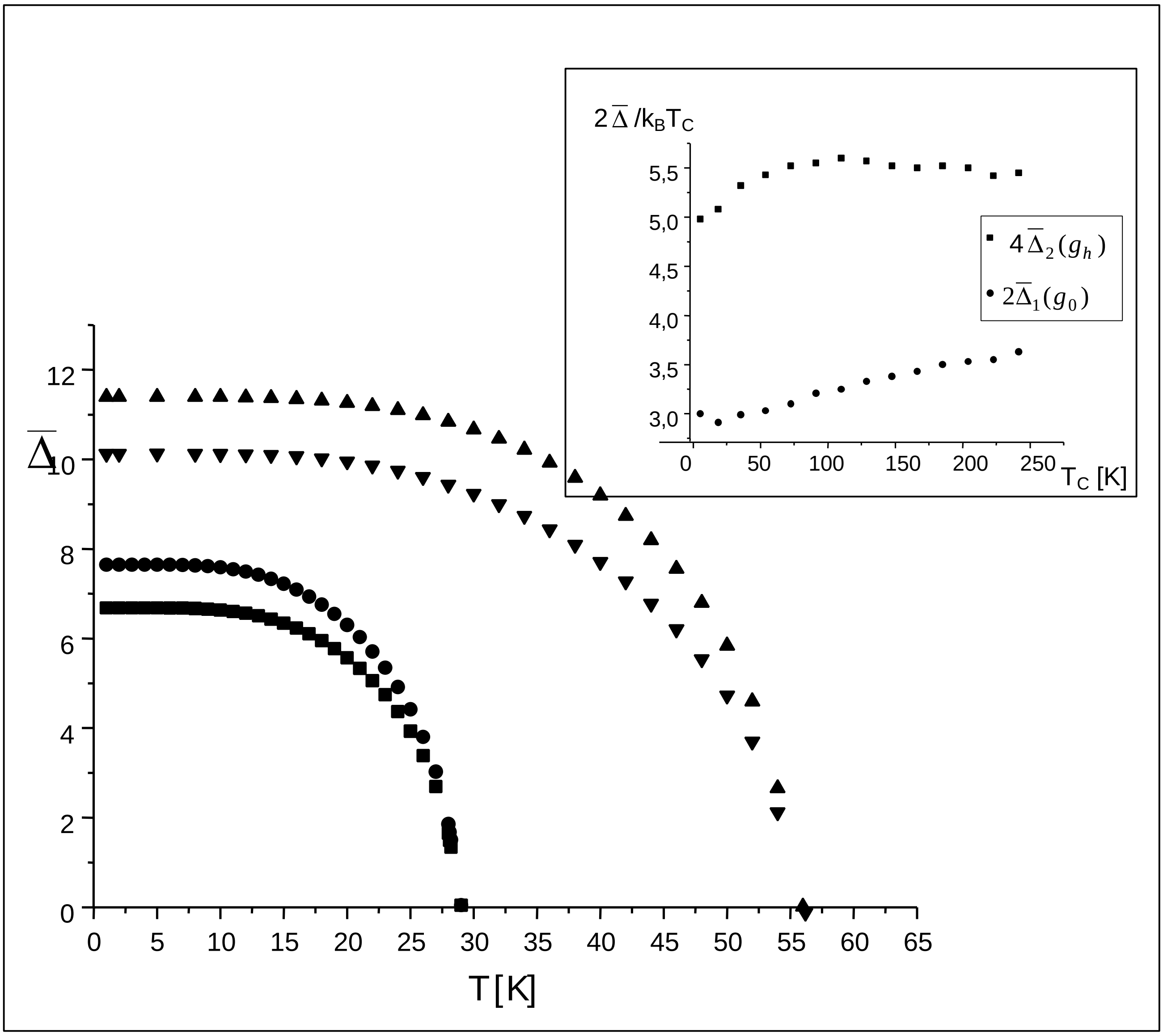

Figure 2: Temperature dependences of the superconducting gaps in meV. Squares and circles refer to $\overline{\Delta}_1(g_0),\overline{\Delta}_2(g_0)$, while down and up triangles are derived for $\overline{\Delta}_2(g_h),\overline{\Delta}_1(g_0)$. The insert shows for the latter case the ratios of the maximal gaps to $T_c$ versus $T_c$ which is equivalent to varying $V_{12}$. The parameters used throughout the paper are given in [17]

again the above two cases, the results shown in figure 3 are obtained. The polaronic band shift $\Delta*$ first increases $T_c$ enormously but then depresses its value to zero with increasing band shift $\Delta^*$. Since the magnitude of $\Delta^*$ depends on the electron-phonon coupling our results show that small and intermediate coupling polaronic distortions lead to a pronounced increase in $T_c$ but reduce $T_c$ in the strong coupling limit. Physically this situation corresponds to an interplay between the interband coupling favouring superconductivity and the lattice distortion which promotes localization. Again a strong enhancement of $T_c$ is observed for the case of two different order parameters as compared to the two s-wave order parameter case.

In conclusion, we have investigated new aspects of the two-band model for superconductivity by considering the influence of different order parameter symmetries on $T_c$ and by studying the effect of a polaronic distortion on it. Combining s and d-wave order parameters always enhances $T_c$ substantially as compared to two isotropic order parameters, since here low energy scales appear from the d-wave channel. The interband coupling also enhances $T_c$ substantially and even at moderate coupling $T_c$ values >100K are obtained. A polaronic distortion favours superconductivity as long as the corresponding electron-phonon interaction is not too large. For intermediate to large values of the coupling, superconductivity is rapidly depressed. The interesting case of the coexistence of superconductivity with a charge density wave instability within the above discussed scenario will be presented elsewhere together with the discussion of the effect of band narrowing on $T_c$.

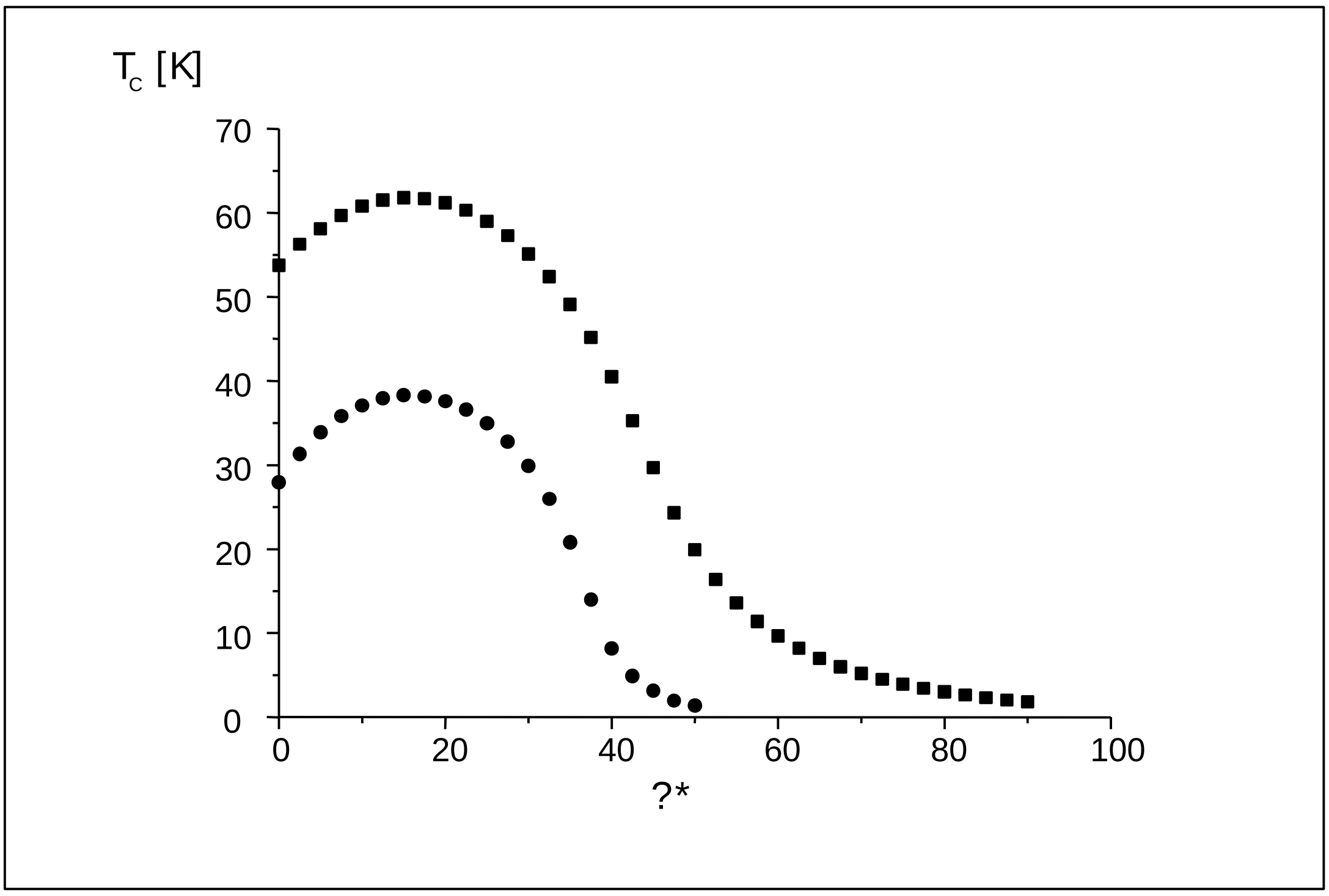

Figure 3: The dependence of $T_c$ on the polaronic shift ?*. The squares refer to s-d coupled order parameters, while the circles correspond to the s-s coupled case.

Acknowledgement: It is a pleasure to acknowledge many useful discussions with H. Büttner, K. A. Müller and A. Simon. Work at Los Alamos is supported by the USDoE. R. M. acknowledges partial support from the State Committee for Scientific Research (KBN Poland): Project No: 2P03B 15422.